\documentclass[twoside]{ilcws07}
\usepackage[latin1]{inputenc}
\usepackage[dvips]{graphicx,epsfig,color}
\usepackage{wrapfig,rotating}
\usepackage{cite,amssymb,amsmath,array,axodraw}

\input paperdef

\begin{document}
\title{
High-Precision Tests of the MSSM with GigaZ} 
\author{S.~Heinemeyer$^1$, W.~Hollik$^2$, A.M.~Weber$^2$ and
G.~Weiglein$^3$
\thanks{email: Georg.Weiglein@durham.ac.uk}
\vspace{.3cm}\\
1- Instituto de Fisica de Cantabria (CSIC-UC), Santander,  Spain\\[.1cm]
2- Max-Planck-Institut f\"ur Physik,
F\"ohringer Ring 6, D--80805 Munich, Germany\\[.1cm]
3- IPPP, University of Durham, Durham DH1~3LE, UK
}

\maketitle

\begin{abstract}
We review the physics potential of the GigaZ option of the International
Linear Collider (ILC) for probing the Minimal Supersymmetric Standard Model
(MSSM) via the sensitivity of the
electroweak precision observables measured at the ILC to quantum
corrections~\cite{slides}. 
A particular focus is put on the effective leptonic weak
mixing angle, $\sweff$. 
The MSSM predictions take into account the complete one-loop results
including the full complex phase dependence, all available MSSM two-loop
corrections as well as the full Standard Model (SM) results. 
We find that the anticipated experimental accuracy at the ILC with GigaZ
option may resolve the virtual effects of SUSY particles even in
scenarios where the SUSY particles are so heavy that they escape direct
detection at the LHC and the first phase of the ILC.
\end{abstract}


\section{Introduction}

Electroweak precision observables (EWPO) are very powerful for testing 
the Standard Model (SM) and extensions of it. A particularly attractive
extension is the Minimal Supersymmetric Standard Model 
(MSSM), see \citere{PomssmRep} for a review of electroweak
precision physics in the MSSM. In this context the $Z$-pole observables
(and also the relation between the $W$- and $Z$-boson masses obtained from
muon decay) play an important role. They comprise in particular
the effective leptonic weak mixing angle, $\sweff$, the total $Z$-boson
width, $\Ga_Z$, the ratio of the hadronic to leptonic decay width of
the $Z$, $R_l$, the ratio of the partial decay width for $Z\to b
\bar{b}$ to the hadronic width,
$R_b$, and the hadronic peak cross section, $\si^0_{\rm had}$. 
Performing fits in constrained SUSY models a certain preference for not
too heavy SUSY particles has been
found~\cite{ehow34,ehoww,AllanachFit,Rotze,mastercode}. 
The prospective improvements in the experimental accuracies, in particular
at the ILC with GigaZ option, will provide a high sensitivity to deviations
both from the SM and the MSSM.
In \refta{tab:expacc} we summarize the current experimental
results~\cite{LEPEWWG,LEPEWWG2,TEVEWWG} together with the anticipated
improvements at the LHC and the ILC with GigaZ option, see
\citeres{blueband,PomssmRep,gigaz,moenig} for details.

\begin{table}[htb!]
\renewcommand{\arraystretch}{1.0}
\BC
\begin{tabular}{|c||c|c|c|c|}
\hline\hline
observable & central exp.\ value & $\si \equiv \si^{\rm today}$ &
             $\si^{\rm LHC}$ & $\si^{\rm ILC/GigaZ}$ \\ \hline \hline
$\MW$ [GeV] & $80.398$ & $0.025$ & $0.015$  & $0.007$ \\ \hline
$\sweff$    & $0.23153$ & $0.00016$ & $0.00020$--$0.00014$ & $0.000013$ 
                                                      \\ \hline
$\Ga_Z$ [GeV] & $2.4952$ & $0.0023$ & --- & 0.001 \\ \hline
$R_l$         & $20.767$ & $0.025$  & --- & 0.01 \\ \hline
$R_b$         & $0.21629$ & $0.00066$ & --- & 0.00014 \\ \hline
$\si^0_{\rm had}$ & $41.540$ & $0.037$ & --- & $0.025$ \\ \hline
$\mt$ [GeV]   & $170.9$  & $1.8$    & $1.0$ & $0.1$ \\
\hline\hline
\end{tabular}
\EC
\renewcommand{\arraystretch}{1}
\caption{Summary of the electroweak precision observables, including the
  top-quark mass, 
their current experimental central values and  experimental errors, 
$\si \equiv \si^{\rm today}$~\cite{LEPEWWG,LEPEWWG2,TEVEWWG}. 
Also shown are the anticipated
experimental accuracies at the LHC, $\si^{\rm LHC}$, and the ILC
(including the GigaZ option), $\si^{\rm ILC}$. Each number represents
the combined results of all detectors and channels at a given collider,
taking into account correlated systematic uncertainties, see
\citeres{blueband,PomssmRep,gigaz,moenig} for details. 
Non-existing analyses are referred to as ``---''.
}
\label{tab:expacc}
\end{table}

In order to confront the predictions of
supersymmetry (SUSY) with the electroweak precision data
and to derive constraints on the supersymmetric parameters,
it is desirable to achieve the same level of accuracy for the
SUSY predictions as for the SM.
In \citeres{MWpope,ZOpope} an new evaluation of $\MW$ and the $Z$-pole
observables in the MSSM has been presented. It includes the full
one-loop result (for the first time with the full complex phase
dependence), all available MSSM two-loop corrections (entering via the
$\rho$~parameter~\cite{dr2lA,drMSSMal2A,drMSSMal2B}), as well as the
full SM results, see \citeres{MWpope, ZOpope} for details. 
The Higgs-boson sector has been implemented including higher-order
corrections (as evaluated with 
{\tt FeynHiggs}~\cite{feynhiggs,mhiggsAEC,mhcMSSMlong}). These
corrections, being formally of higher-order, can give sizable
contributions to the EWPO. 
The remaining
theory uncertainties have been estimated to be 
$\de\MW^{\rm theo} \lsim 10 \mev$~\cite{MWpope} and 
$\de\sweff^{\rm theo} \lsim 7 \times 10^{-5}$~\cite{ZOpope}.
It has furthermore been shown in \citere{ZOpope} that $\MW$, $\sweff$
and $\Ga_Z$ show a pronounced sensitivity to the SUSY parameters,
while the other EWPO exhibit only a small variation over the MSSM
parameter space. In view of the extraordinary anticipated accuracy of
$\de\sweff^{\rm ILC/GigaZ} = 1.3 \times 10^{-5}$~\cite{moenig}, the
effective leptonic weak mixing angle will be a highly sensitive probe of
electroweak physics.


\section{\boldmath{$\sweff$} in a global MSSM scan}

We first analyse the sensitivity of $\sweff$ to higher-order effects in
the MSSM by
scanning over a broad range of the SUSY parameter space. The following SUSY
parameters are varied independently of each other in a random parameter scan
within the given range:
\begin{eqnarray}
 {\rm sleptons} &:& M_{{\tilde F},{\tilde F'}}= 100\dots2000\gev, \non \\
 {\rm light~squarks} &:& M_{{\tilde F},{\tilde F'}_{\textup{up/down}}}
                   = 100\dots2000\gev, \non \\
 \Stop/\Sbot {\rm ~doublet} &:& 
                         M_{{\tilde F},{\tilde F'}_{\textup{up/down}}}
                    = 100\dots2000\gev, 
 \quad A_{\tau,t,b} = -2000\dots2000\gev, \non \\
 {\rm gauginos} &:& M_{1,2}=100\dots2000\gev, 
 \quad \mgl=195\dots1500\gev, \non \\
 && \mu = -2000\dots2000\gev,\non \\ 
 {\rm Higgs} &:& \MA=90\dots1000\gev, 
 \quad\tb = 1.1\dots60.
\label{scaninput}
\end{eqnarray}
Here $M_{{\tilde F},{\tilde F'}}$ are the diagonal soft SUSY-breaking
parameters in the sfermion sector, $A_f$ denote the trilinear couplings,
$M_{1,2}$ are the soft SUSY-breaking parameters in the
chargino and neutralino sectors, $\mgl$ is the gluino mass, $\mu$ the Higgs
mixing parameter, $\MA$ the $\cp$-odd Higgs boson mass, and $\tb$ is the
ratio of the two vacuum expectation values.
Only the constraints on the MSSM parameter space
from the LEP Higgs searches~\cite{LEPHiggsMSSM,LEPHiggsSM} and the lower
bounds on the SUSY particle masses from direct searches as given 
in \citere{pdg} were taken into account.
Apart from these constraints no other restrictions on the MSSM parameter
space were made.

\begin{figure}[htb!]
\begin{center}
\includegraphics[width=12.7cm,height=8cm]{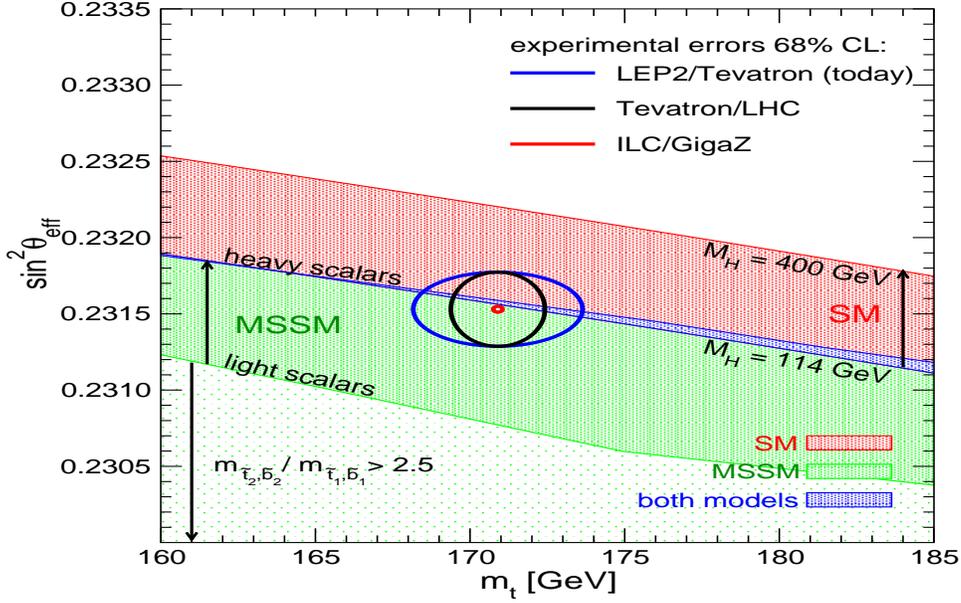}
\begin{picture}(0,0)
\CBox(-290,190)(-195,220){White}{White}
\end{picture}
\end{center}
\vspace{-2em}
\caption{MSSM parameter scan for $\sweff$ as a function of $\mt$ over the
  ranges given in \refeq{scaninput}.  Todays 68\%~C.L.\ ellipses
  as well as future precisions, drawn around todays central value,  are
  indicated in the plot.}  
\label{fig:Scans2} 
\end{figure}

In \reffi{fig:Scans2} we
compare the SM and the MSSM predictions for $\sweff$
as a function of $\mt$ as obtained from the scatter data. 
The predictions within the two models 
give rise to two bands in the $\mt$--$\sweff$ plane with only a relatively 
small overlap region (indicated by a dark-shaded (blue) area).
The allowed parameter region in the SM (the medium-shaded (red)
and dark-shaded (blue) bands) arises from varying the only free parameter 
of the model, the mass of the SM Higgs boson, from $\MHSM = 114\gev$, the LEP 
exclusion bound~\cite{LEPHiggsSM}
(lower edge of the dark-shaded (blue) area), to $400 \gev$ (upper edge of the
medium-shaded (red) area).
The very light-shaded (green), the light shaded (green) and the
dark-shaded (blue) areas indicate allowed regions for the unconstrained
MSSM. In the very light-shaded region at least one of the ratios
$\mstz/\mste$ or $\msbz/\msbe$ exceeds~2.5 (with the 
convention that $\msfe \le \msfz$), 
while the decoupling limit with SUSY masses of \order{2 \tev}
yields the upper edge of the dark-shaded (blue) area. Thus, the overlap 
region between the predictions of the two models corresponds in the SM
to the region where the Higgs boson is light, i.e., in the MSSM allowed
region ($\Mh \lsim 130 \gev$~\cite{feynhiggs,mhiggsAEC}). In the MSSM it
corresponds to the case where all 
superpartners are heavy, i.e., the decoupling region of the MSSM.
The 68\%~C.L.\ experimental results
for $\mt$ and $\sweff$ are indicated in the plot. As can be seen from
\reffi{fig:Scans2}, the current 
experimental 68\%~C.L.\ region for 
$\mt$ and $\sweff$ is in good agreement with both models and does not 
indicate a preference for one of the two models.
The prospective accuracies for the Tevatron/LHC 
and the ILC with GigaZ option, see \refta{tab:expacc}, 
are also shown in the plot (using the current central values).
Especially the ILC/GigaZ precision indicates the strong
potential for a significant improvement of the
sensitivity of the electroweak precision tests~\cite{gigaz}.
A comparison of the MSSM parameter space preferred by $\sweff$ and the
directly measured values will constitute a highly sensitive test of
the model.


\section{Scenario where no SUSY particles are observed 
     at the LHC} 
\label{sec:ILCscen}

It is interesting to investigate whether the high accuracy achievable at
the GigaZ option of the ILC would provide sensitivity to indirect effects of
SUSY particles even in a scenario where the (strongly interacting) 
superpartners are so heavy that they escape detection at the LHC.

\begin{figure}[bth!]
\begin{center}
\includegraphics[width=12.7cm,height=8cm]{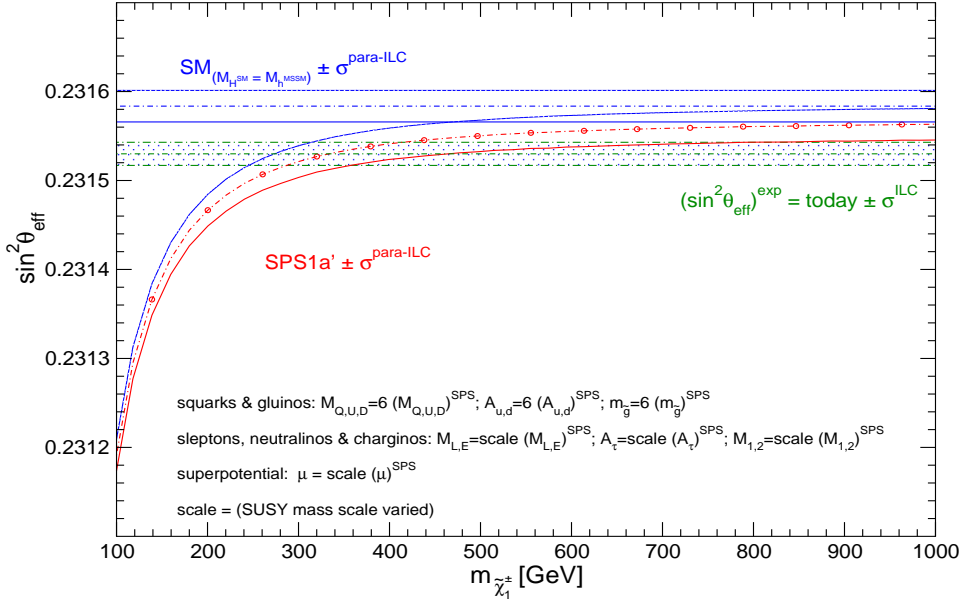}
\vspace{-1.0em}
\caption{
Theoretical prediction for $\sweff$ in the SM and the MSSM (including
prospective parametric theoretical uncertainties) compared to
the experimental precision at the ILC with GigaZ option.  
An SPS1a$'$ inspired scenario is used, where the squark and gluino
mass parameters
are fixed to 6~times their SPS~1a$'$ values. The other mass 
parameters are varied with a common scalefactor.}
\label{fig:ILC} 
\end{center}
\end{figure}

We consider in this context a scenario with very heavy squarks and a 
very heavy gluino. It is based on the values of the SPS~1a$'$ benchmark
scenario~\cite{sps}, but the squark and gluino
mass parameters
are fixed to 6~times their SPS~1a$'$ values. The other masses are 
scaled with a common scale factor 
except $\MA$ which we keep fixed at its SPS~1a$'$ value.
In this scenario 
the strongly interacting particles are too heavy to be detected at the
LHC, while, depending on the scale-factor, some colour-neutral particles
may be in the ILC reach. In \reffi{fig:ILC} we show the prediction for
$\sweff$ in
this SPS~1a$'$ inspired scenario as a function of the lighter chargino
mass, $\mcha{1}$. The prediction includes the parametric
uncertainty, $\si^{\rm para-ILC}$, induced by the ILC measurement of $\mt$, 
$\de\mt = 100 \mev$~\cite{mtdet12}, and the numerically more
relevant prospective future uncertainty on $\De\al^{(5)}_{\textup{had}}$,
$\de(\De\al^{(5)}_{\textup{had}})=5\times10^{-5}$~\cite{fredl}. 
The MSSM prediction for $\sweff$
is compared with the experimental resolution with GigaZ precision,
$\si^{\rm ILC} = 0.000013$, using for simplicity the current
experimental central value. The SM prediction (with 
$\MHSM = \Mh^{\rm MSSM}$) is also shown, applying again the parametric 
uncertainty $\si^{\rm para-ILC}$.

Despite the fact that no coloured SUSY 
particles would be observed at the LHC in this scenario, the ILC with
its high-precision 
measurement of $\sweff$ in the GigaZ mode could resolve indirect effects
of SUSY up to $m_{\tilde\chi^\pm_1} \lsim 500 \gev$. This means that the
high-precision measurements at the ILC with GigaZ option could be
sensitive to indirect effects of SUSY even in a scenario where SUSY
particles have {\em neither \/} been directly detected at the LHC nor the
first phase of the ILC with a centre of mass energy of up to $500 \gev$.


\section{Conclusions}

EWPO provide a very powerful test of the SM
and the MSSM. We have reviewed results for $\MW$ and $Z$~boson
observables such as $\sweff$, $\Ga_Z$, 
$R_{\rm l}$, $R_{\rm b}$, $\sigma_{\textup{had}}^0$. Within the MSSM 
new results for the EWPO containing the complete one-loop results with
complex parameters and all available higher-order corrections in the SM
and the MSSM have recently become available.
The sensitivity to higher-order effects will drastically improve with
the ILC precision (including the GigaZ option) on the EWPO and
$\mt$. This has been illustrated in two examples. A general scan over
the MSSM parameter space for $\sweff$ and $\mt$ currently does not
prefer the SM or the MSSM over the other. However, the anticipated
GigaZ precision indicates the high 
potential for a significant improvement of the
sensitivity of the electroweak precision tests.
In a second example we have assumed a scenario with very heavy SUSY
particles, outside the reach of the LHC and the first stage of the ILC
with $\sqrt{s} = 500 \gev$. It has been shown that even in such a
scenario the GigaZ precision on $\sweff$ may resolve virtual effects of
SUSY particles, providing a possible hint to the existence of new physics.


\bigskip
\noindent
{\bf Acknowledgements}

\noindent
We thank G.~Moortgat-Pick for interesting discussions concerning 
\refse{sec:ILCscen}. 
Work supported in part by the European Community's Marie-Curie Research
Training Network under contract MRTN-CT-2006-035505
`Tools and Precision Calculations for Physics Discoveries at Colliders'
(HEPTOOLS). 


\begin{footnotesize}

\end{footnotesize}

\end{document}